\documentclass[11pt]{article}
\usepackage{geometry}                
\usepackage{graphicx}
\usepackage{rawfonts}
\usepackage{amssymb}
\usepackage{epstopdf}
\usepackage{amsmath,amsthm,amssymb}
\usepackage{mathtools}
\usepackage{fullpage}
\usepackage{color}
\usepackage[usenames,dvipsnames,svgnames]{xcolor}
\usepackage[colorlinks=true,citecolor=blue,linkcolor=red,urlcolor=blue]{hyperref}
\usepackage[font=footnotesize, labelfont=bf, margin=0.5cm]{caption}
\usepackage[labelformat=simple]{subcaption}
\usepackage{rawfonts}
\usepackage{etex}

\usepackage{ulem}

\newtheorem{theo}{Theorem}

\newtheorem{remark}{Remark}

\def\Pr{\noindent \textit{Proof: }}
\def\qed{$\Box$}

\usepackage{tikz}

\def\2;{\;\;}

\title{Lattice tadpoles}

\author
{S.~G.~Whittington\thanks{\href{mailto:swhittin@chem.utoronto.ca}{stuart.whittington@utoronto.ca}}\\
\small Department of Chemistry\\
\small University of Toronto, Toronto, Ontario M5S 3H6, Canada
}

\begin{document}

\maketitle

\begin{abstract}
We prove several rigorous results about the asymptotic behaviour of the numbers of tadpoles (or lassos) embedded in a lattice, including cases where the head is subject to a constraint like being unknotted, or where the tail pierces the surface spanned by the head.  Similar results can be proved for other homeomorphism types such as dumbbells, twin tailed tadpoles and two tailed tadpoles.
\end{abstract}

\maketitle

\section{Introduction}
\label{sec:Introduction}
A \textit{tadpole} is a connected graph with one vertex of degree 1 and one vertex of degree 3.  All other vertices have degree 2, and the cyclomatic index is 1.  The cycle is often called the \textit{head} of the tadpole and the walk between the vertices of degree 1 and 3 is called the \textit{tail}.    Tadpoles have been studied in molecular biology, where they are called \textit{lassos} and occur as a motif in protein structure \cite{Sulkowska2019, Sulkowska2016}.  More complicated homeomorphism types, such as those related to $\theta$-graphs \cite{Sulkowska2024}, have also been observed in proteins.  

Tadpoles have been investigated in statistical mechanics, as a contributing graph type in Sykes' counting theorem for self-avoiding walks \cite{Sykes} and in the limiting ring closure probability for self-avoiding walks \cite{GuttmannSykes,WhittingtonTruemanWilker}.  They also occur in the high temperature expansion of the susceptibility of the Ising model \cite{Domb,Stanley}.  More complex structures, including $\theta$-graphs \cite{BeatonOwczarek}, have been investigated numerically.

When tadpoles appear in protein structures, the interest centres on various particular structure types that can occur \cite{Sulkowska2019,Sulkowska2016}.  The head spans a minimal surface and the tail can pierce this surface one or more times.  The head can have a specific knot type and the surface spanned by the head is then the Seifert surface of the knot.  Again, the tail can pierce this surface.

\begin{figure}
\centering
\includegraphics[scale=0.3]{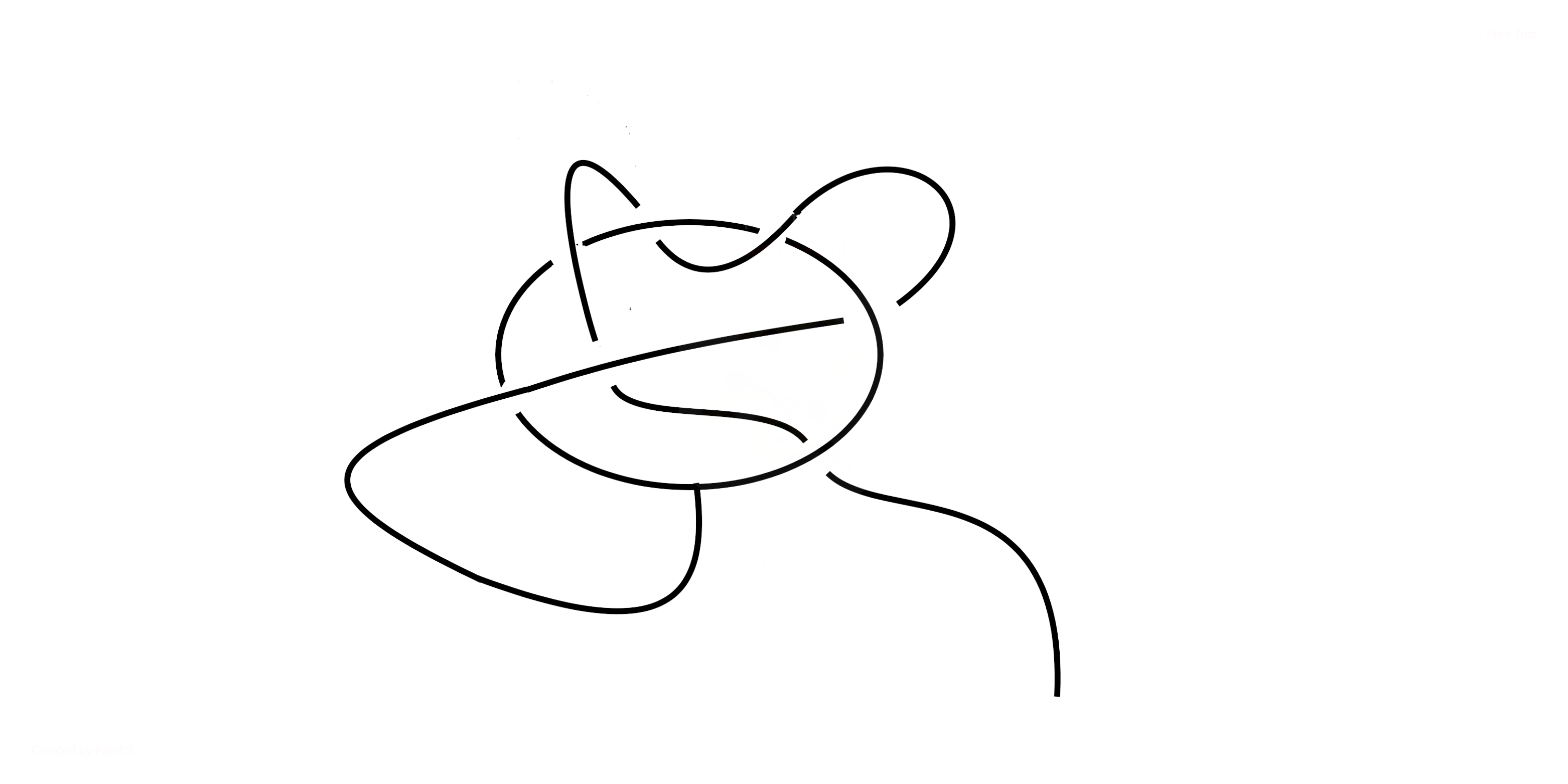}
\caption{A tadpole with unknotted head where the tail pierces the head three times.}
\label{fig:piercedhead}
\end{figure}

In this paper we investigate the asymptotics of the numbers of tadpoles, embedded in a lattice, in Section \ref{sec:tadpoles}, where we consider all tadpoles with a fixed total number, $n$,  of edges, or tadpoles with $h$ edges in the head and $n-h$ edges in the tail.  We shall be interested in cases where both $h$ and $n$ go to infinity and where $n$ goes to infinity with $h$ fixed.  We shall also be interested in cases where the head is unknotted or has a particular knot type, and where the tail pierces the head a given number of times.  Figure \ref{fig:piercedhead} shows a tadpole with unknotted head, where the head is pierced three times by the tail.
In Section \ref{sec:others} we give some similar results for other homeomorphism types, such as dumbbells,  which also occur as a contributing graph type in Sykes' counting theorem \cite{Sykes} and as a motif in protein structures \cite{Sulkowska2019}.  We conclude with a short Discussion in Section \ref{sec:discussion}.

\section{Some preliminary results}
\label{sec:preliminary}

We shall need several definitions and results that are already in the literature and we shall collect them in this section for convenience.  We shall primarily be concerned with the $d$-dimensional hypercubic lattice, $Z^d$, and especially with the square and simple cubic lattices.  Suppose that $c_n$ is the number of self-avoiding walks with $n$ edges in $Z^d$.  Then \cite{Hammersley1957}
\begin{equation}
\log d \le \lim_{n\to\infty} n^{-1} \log c_n = \inf_{n > 0} n^{-1} \log c_n = \kappa_d \le \log(2d-1)
\end{equation}
where $\kappa_d$ is the \textit{connective constant} of the lattice.  If $p_n$ is the number of (undirected, unrooted) polygons in $Z^d$ then \cite{Hammersley1961}
\begin{equation}
\lim_{n\to\infty} n^{-1} \log p_n = \sup_{n > 0} n^{-1} \log p_n = \kappa_d
\label{eqn:polygon}
\end{equation}
so that polygons and self-avoiding walks have the same exponential growth rate.

We shall require two results about \textit{patterns} in self-avoiding walks and polygons.  A pattern is any fixed self-avoiding walk and a \textit{Kesten pattern} is a pattern that occurs three times on at least one self-avoiding walk.  Suppose that $P$ is a Kesten pattern and let $c_n(\bar{P})$ be the number of $n$-edge self-avoiding walks in which $P$ never occurs.  Then Kesten \cite{Kesten} showed that 
\begin{equation}
\limsup_{n\to\infty} n^{-1} \log c_n(\bar{P}) < \kappa_d
\label{eqn:Kesten}
\end{equation}
so that walks not containing $P$ are exponentially rare.

A special kind of pattern is an \textit{end pattern}.  An end pattern is a pattern that occurs at either the beginning or end of a self-avoiding walk that can be extended indefinitely.  If $P$ is an end pattern, and $c_n(P)$ is the number of $n$-edge self-avoiding walks that have $P$ as the first or last few edges, then Madras \cite{Madras} showed that
\begin{equation}
\liminf_{n\to\infty} \frac{c_n(P)}{c_n} > 0.
\label{eqn:endpattern}
\end{equation}

We shall also require some results about the unknot probability for polygons in $Z^3$.  Suppose that $p_n^0$ is the number of polygons in $Z^3$ with $n$ edges, that are unknotted.  Then \cite{Pippenger, SumnersWhittington}
\begin{equation}
\lim_{n \to \infty} n^{-1} \log p_n^0 = \kappa_3^0 < \kappa_3
\label{eqn:unkotting}
\end{equation}
so unknotted polygons are exponentially rare in the set of all polygons.

Finally, a \textit{positive walk} is a self-avoiding walk that starts in, and is is restricted to lie in or on one side of, a lattice hyperplane.  If $c_n^+$ is the number of $n$-edge positive walks then \cite{Whittington}
\begin{equation}
\lim_{n \to \infty} n^{-1} \log c_n^+ = \kappa_d.
\label{eqn:positive}
\end{equation}

\section{Tadpoles}
\label{sec:tadpoles}

In this section we prove several rigorous results about the asymptotics of the numbers of tadpoles, with and without various constraints.  Suppose that 
$t(h,n-h)$ is the number of tadpoles with $h$ edges in the head and $n-h$ in the tail. The total number of tadpoles with $n$ edges is then
\begin{equation}
t_n = \sum_h t(h,n-h).
\end{equation}

\subsection{Tadpoles with a fixed total number of edges}
\label{sec:alltadpoles}

\begin{theo}
If $t_n$ is the number of tadpoles with $n$ edges and $c_n$ is the number of self-avoiding walks with $n$ edges then there are two positive numbers, $A_1$ and $A_2$ such that
$$ 0 < A_1 \le \liminf_{n \to \infty} \frac{t_n}{c_n} \le \limsup_{n \to \infty} \frac{t_n}{c_n} \le  A_2 \le 1. $$
\label{thm:Thm1}
\end{theo}

\Pr
To get an upper bound we delete an edge in the head of the tadpole, incident on the vertex of degree 3, to give a self-avoiding walk with one less edge.  Hence $t_n \le c_{n-1} \le c_n$ where we have used the fact that $c_{n-1} \le c_n$ \cite{O'Brien}.  For the lower bound we use Madras' end pattern theorem \cite{Madras}.  Suppose that $P_0$ is a pattern consisting of three edges in a U-shape, so that $P_0$ can be converted to a unit square by adding an edge.The number of self-avoiding walks with $n-1$ edges having $P_0$ as the starting end pattern is $c_{n-1}(P_0)$ and 
$\liminf_{n \to \infty} c_{n-1}(P_0)/c_{n-1} > 0$ \cite{Madras}.  Each of the walks containing this end pattern can be converted to a tadpole with a head of size 4, giving the string of inequalities:
$$\liminf_{n \to \infty} \frac{t_n}{c_n} \ge \liminf_{n \to \infty} \frac{c_{n-1}(P_0)}{c_n} \ge \liminf_{n \to \infty} \frac{c_{n-1}(P_0)}{(2d-1)c_{n-1}} \ge A_1 > 0.$$
This completes the proof.
\qed

It is believed that 
\begin{equation}
c_n = B n^{\gamma -1} \mu_d^n (1+o(1))
\end{equation}
where $\mu_d = e^{\kappa_d}$.  $\gamma $ is a critical exponent.  If $c_n$ has this form, that is if the critical exponent exists, then
Theorem \ref{thm:Thm1} shows that tadpoles and walks have the same critical exponent.  This was proved in 1978 \cite{GuttmannWhittington} by a completely different argument based on Sykes' counting theorem \cite{Sykes}.

If the head and tail sizes diverge together, we are interested in $t(\alpha n, (1 - \alpha)n)$ for $0 < \alpha < 1$, and we have the following theorem:
\begin{theo}
For $0 < \alpha < 1$ 
$$\lim_{n \to \infty} n^{-1} \log t(\alpha n, (1 - \alpha) n) = \kappa_d$$
\label{theo:divergetogether}
\end{theo}
\Pr
If we delete an edge in the head, incident on the vertex of degree 3 we obtain a self-avoiding walk with $n-1$ edges, so 

$$t(\alpha n, (1 - \alpha)n) \le c_{n-1}$$
and $\limsup_{n \to \infty} n^{-1} \log t(\alpha n, (1 - \alpha)n) \le \kappa_d$.
We can obtain a lower bound by concatenating a polygon and a positive walk by translating to that the distinguished degree 1 vertex of the positive walk coincides with the top vertex of the polygon.  This gives the bound
$$t(\alpha n, (1 - \alpha)n) \ge p_{\alpha n} c_{(1-\alpha)n}$$
and $ \liminf_{n \to \infty} n^{-1} \log 
t(\alpha n, (1 - \alpha)n) \ge \kappa_d$.  This completes the proof.
\qed

\begin{remark}
This result also follows as a special case of Lemma 4.6 in \cite{SoterosSumnersWhittington}.
\end{remark}

\begin{figure}
\centering
\includegraphics[scale=0.4]{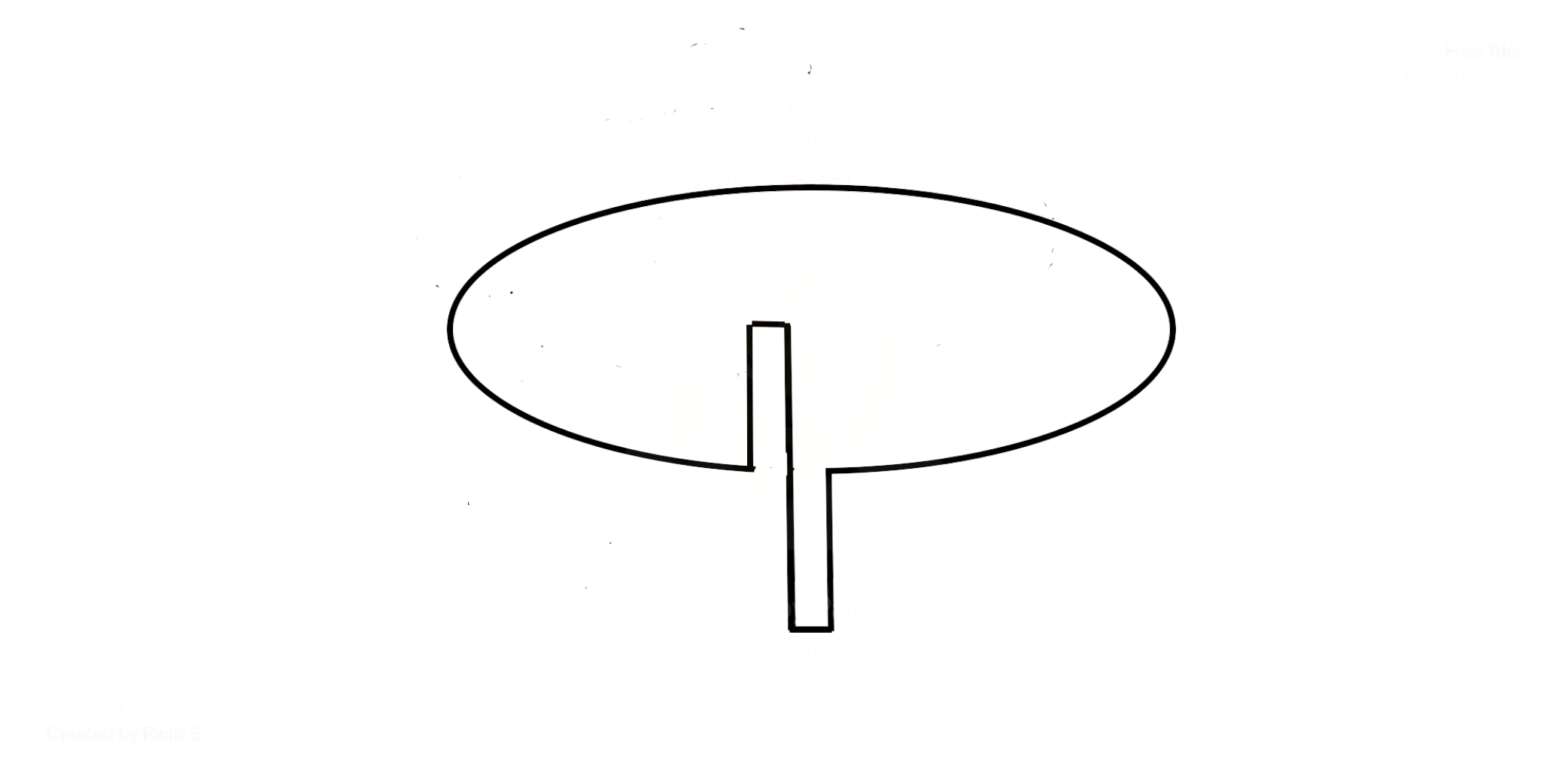}
\caption{A polygon in two dimensions containing the pattern $P_1$.}
\label{fig:polypattern}
\end{figure}

\subsection{Tadpoles with the tail inside the head}
\label{sec:reentrant}

\begin{figure}
\centering
\includegraphics[scale=0.4]{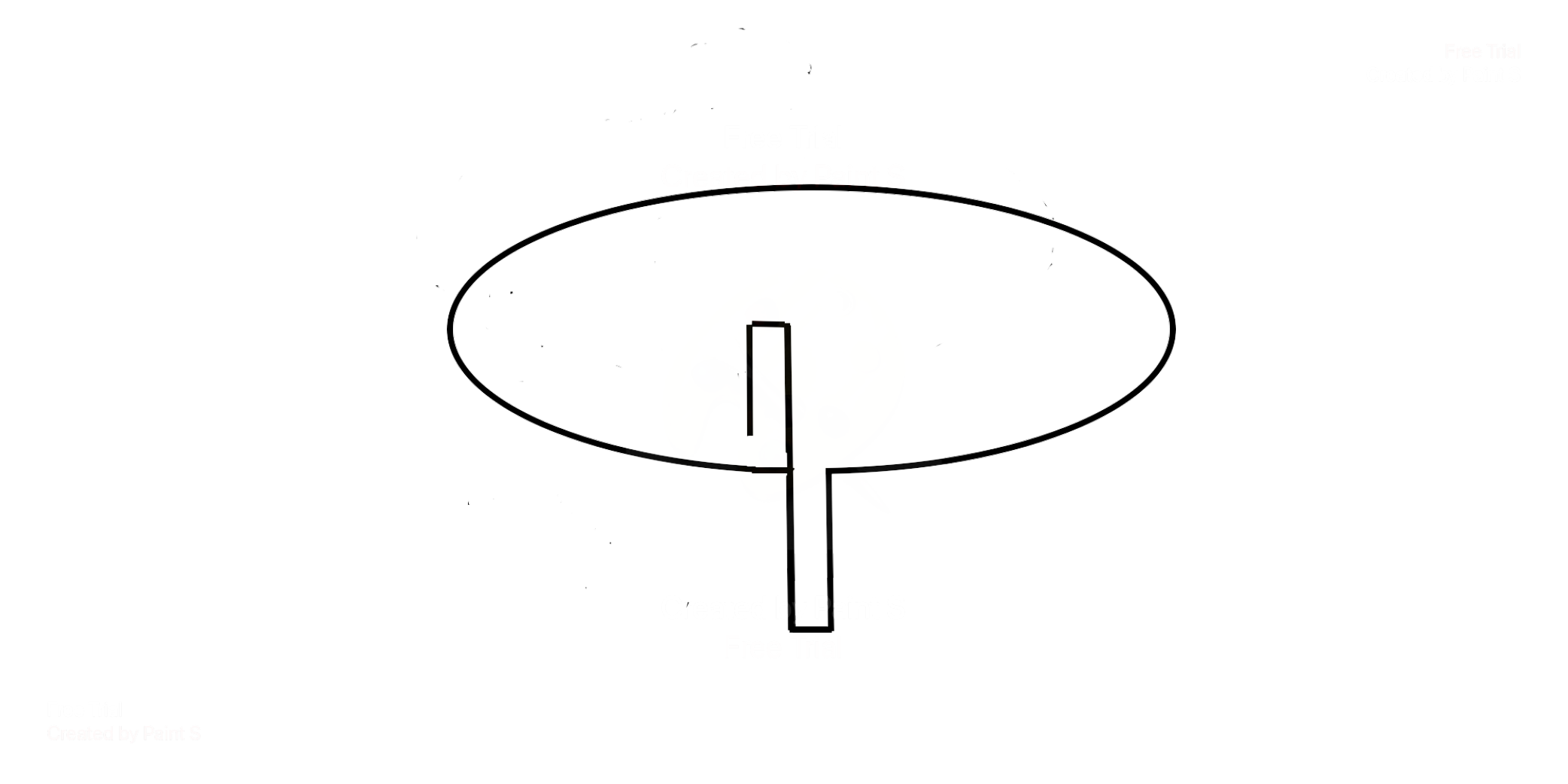}
\caption{A tadpole obtained by modifying the pattern $P_1$.}
\label{fig:tadreentrant}
\end{figure}

In two dimensions the tail of the tadpole can be \textit{inside} the head, and we examine this situation in this Section.  We first show that tadpoles with this constraint have the same exponential growth rate as all tadpoles.  Suppose that the number of tadpoles in the square lattice with $n$ edges, with the tail inside the head, is $\hat{t}_n$.  We have the following theorem:

\begin{theo}
For $\hat{t}_n$, the number of tadpoles on the square lattice with $n$ edges, with the tail inside the head
$$\lim_{n\to \infty} n^{-1} \log \hat{t}_n = \kappa_2.$$
\label{theo:tadpolereentrant}
\end{theo}
\Pr 
To obtain an upper bound on $\hat{t}_n$ we observe that $\hat{t}_n \le t_n = e^{\kappa_2 n + o(n)}$.The lower bound comes from Kesten's pattern theorem \cite{Kesten} adapted to work for polygons \cite{SumnersWhittington}.  Suppose that $m$ is even. Define the pattern $P_1$ as follows.  It consists of $2(m+1)$ edges: 
$$P_1 = (0,0)-(0,1)- \dots -(0,m/2)-(1,m/2)-(1,m/2 -1)- \ldots $$
$$ -(1,-m/2)-(2,-m/2)-(2,-m/2+1) -...-(2,0).$$
Figure \ref{fig:polypattern} is a sketch of a polygon containing $P_1$.
$P_1$ occurs three times on at least one self-avoiding walk so $P_1$ is a Kesten pattern and it occurs on all except exponentially few sufficiently long polygons.  For the proof of this theorem we could take $m=2$.  If $p_n(\bar{P}_1)$ is the number of polygons on which $P_1$ does not occur than $\limsup_{n\to\infty}n^{-1} \log p_n(\bar{P}_1)
< \kappa_2$. We can convert a polygon containing $P_1$ into a tadpole by either (a) deleting the edge $(0,0)-(0,1)$ and adding the edge $(0,0)-(1,0)$ or (b) by deleting the edge $(2,0)-(2,-1)$ and adding the edge $(1,0)-(2,0)$.  See Figure \ref{fig:tadreentrant} for a sketch of case (a).  Each construction gives a tadpole and one gives a tadpole with the tail inside the head.  Choose this construction.  Then
\begin{equation}
\hat{t}_n \ge p_n \left(1-e^{-\beta n}\right)
\end{equation}
for some $\beta > 0$.  Hence $\liminf_{n\to\infty} n^{-1} \log \hat{t}_n \ge \kappa_2$.  This and the upper bound given above complete the proof.
\qed

If the tail is small compared to the head, we can prove a stronger result.  If $\hat{t}(n-m,m)$ is the number of tadpoles in the square lattice with head of size $n-m$ and tail of size $m$, with the tail inside the head, then

\begin{theo}
With $m$ fixed we have 
$$\epsilon n p_n \left(1- e^{-\beta n}\right) \le \hat{t}(n-m,m) \le (n-m) p_{n-m} c_m$$
for some $\epsilon > 0$ and some $\beta > 0$, 
so that the asymptotics are those of a rooted polygon.
\end{theo}
\Pr 
To get an upper bound consider adding a self-avoiding walk with $m$ edges at each vertex of a polygon with $n-m$ edges.  The resulting graphs include tadpoles with $n-m$ edges in the head and $m$ edges in the tail so
$$\hat{t}(n-m,m) \le (n-m) p_{n-m} c_m.$$
The lower bound comes from an argument that is a modification and extension of the argument used in the proof of Theorem \ref{theo:tadpolereentrant}.  If $m$ is even the pattern $P_1$ will occur at least $\epsilon n$ times in all except exponentially few polygons with $n$ edges, for some $\epsilon > 0$ \cite{Kesten}.  We can choose one of these patterns in $\epsilon n$ ways, and modify it to form a tadpole with the tail inside the head (see Figure \ref{fig:tadreentrant}), so that
$$\hat{t}(n-m,m) \ge \epsilon n p_n \left(1-e^{-\beta n}\right)$$
for some $\beta > 0$.
If $m$ is odd and $m \ge 3$ we can construct a pattern $P_2$ as follows.  Let $q=(m+1)/2$.  Then
$$P_2 = (0,0)-(0,1)- \ldots -(0,q)-(1,q)-(1,q-1)-\ldots -(1,-q)-(2,-q)- \ldots -(2,0)$$
and the rest of the argument follows \textit{mutatis mutandis} except that we delete two consecutive edges, instead of one.  A similar argument can be constructed for $m=1$.
\qed

\begin{remark}
Using the other version of the construction for the lower bound proves the same result for tadpoles in $Z^2$ with the short tail outside the head.  A similar but simpler argument provides a proof that all tadpoles with a short tail and large head in $Z^d$ have the same asymptotics as  rooted polygons.
\end{remark}

\begin{remark}
Similar pattern theorem arguments can be used to prove that figure eights with one small and one large cycle, and $\theta$-graphs with two short branches and one long branch have the same asymptotics as  rooted polygons.  
\end{remark}

\subsection{Tadpoles in $Z^3$  with constraints}
\label{sec:constraints}

In this section we concentrate on the case where $d=3$ and we consider the effects of some other constraints, such as fixing the knot type of the head, or fixing the number of times that the tail pierces the minimal surface spanned by the head.

Suppose that $t_{K}(h,n-h)$ is the number of tadpoles in $Z^3$ with $h$ edges in the head and $n-h$ edges in the tail, with the head having knot type K.  For the special case that $K$ is the unknot $K=\emptyset$ we have the following theorem:
\begin{theo}
For $0 < \alpha < 1$
$$\lim_{n\to\infty} n^{-1} \log t_{\emptyset}(\alpha n, (1-\alpha) n) = \alpha \kappa_3^0 +(1-\alpha) \kappa_3.$$
\label{theo:unknottedhead}
\end{theo}
\Pr 
The lower bound comes from concatenating an unknotted polygon and a positive walk, as in the proof of Theorem \ref{theo:divergetogether}, giving
$$t_{\emptyset}(\alpha n, (1-\alpha) n) \ge p^0_{\alpha n} c^+_{(1-\alpha) n}$$
To obtain an upper bound we concatenate an unknotted polygon and a self-avoiding walk at every vertex of the polygon, giving
$$t_{\emptyset}(\alpha n, (1-\alpha) n) \le \alpha n p^0_{\alpha n} c_{(1-\alpha)n}.$$
Taking logarithms in the lower and upper bounds, dividing by $n$ and letting $ \to \infty$ completes the proof.
\qed

The probability that the head is unknotted, $Prob (unknot)$, goes to zero like
\begin{equation}
Prob (unknot) = t_{\emptyset}(\alpha n, (1-\alpha)n)/ t (\alpha n, (1-\alpha)n)= \exp[-\alpha (\kappa_3 -\kappa_3^0)n + o(n)]
\end{equation}
for $0 < \alpha < 1$, where $\kappa_3^0 < \kappa_3$ \cite{Pippenger, SumnersWhittington}.  For a related result in a more general setting, see \cite{SoterosSumnersWhittington}, Theorem 4.8.

If $K$ is any knot type other than the unknot, then:

\begin{theo}
For $0 < \alpha < 1$
$$\alpha \kappa_3^0 + (1-\alpha) \kappa_3 \le \liminf_{n\to\infty} n^{-1} \log t_{K}(\alpha n, (1-\alpha) n)
\le \limsup_{n \to \infty} n^{-1} \log t_K(\alpha n, (1-\alpha) n) <   \kappa_3.$$
\label{theo:knottedhead}
\end{theo}
\Pr 
This follows from an argument similar to that used in the proof of Theorem \ref{theo:unknottedhead}.
\qed

This shows that, when $0 < \alpha < 1$, each individual knot type for the head is exponentially rare.

\begin{figure}
\centering
\includegraphics[scale=0.4]{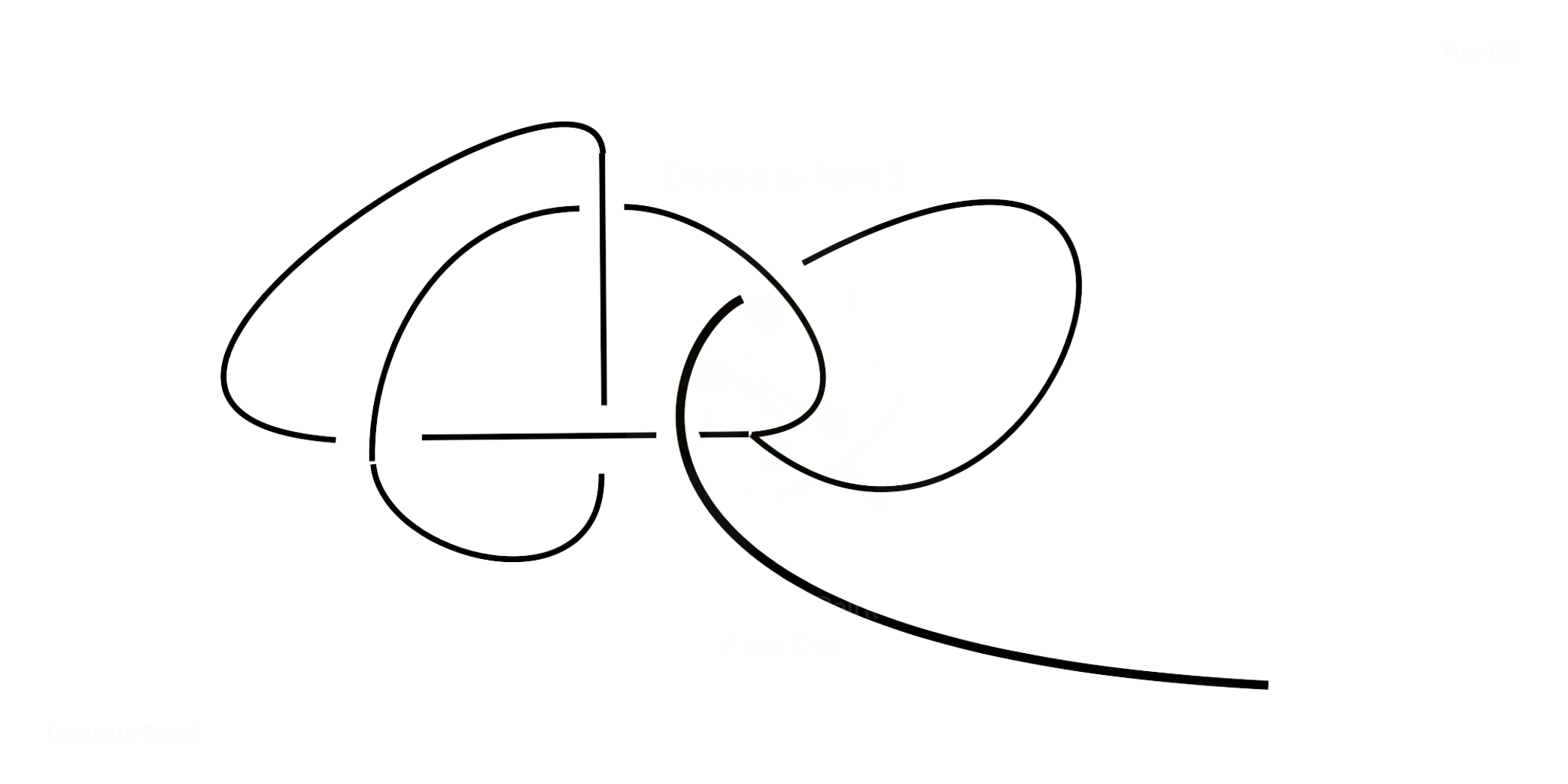}
\caption{A tadpole where the head is a trefoil and the tail pierces the head once.}
\label{fig:trefoilpierced}
\end{figure}

A different constraint, which is of interest in the molecular biology of proteins \cite{Sulkowska2019, Sulkowska2016}, is the number of times that the head is pierced by the tail.  In three dimensions, the head spans a minimal surface and, if the head is knotted, this surface is the Seifert surface of the knot.  One can ask for the number of times that the tail pierces this surface.  We write $t^{[k]}(h,n-h)$ for the number of tadpoles in $Z^3$ with head of size $h$ and tail of size $n-h$ where the tail pierces the surface spanned by the head exactly $k$ times.  If we want to restrict to a particular knot type for the head, $K$ say, we write $t_{K}^{[k]}(h,n-h)$ for the corresponding count.  Figure \ref{fig:trefoilpierced} shows a tadpole with a trefoil head pierced once by the tail.

We have the following theorem:

\begin{theo}
If $0 < \alpha < 1$ and the head is pierced exactly $k$ times by the tail, then
$$\lim_{n \to \infty} n^{-1} \log t^{[k]}(\alpha n, (1-\alpha)n) = \kappa_3.$$
For the special case where the head is unknotted
$$\lim_{n \to \infty} n^{-1} \log t_{\emptyset}^{[k]}(\alpha n, (1-\alpha)n) = \alpha \kappa_3^0 + (1-\alpha) \kappa_3.$$
\label{theo:piercings}
\end{theo}
\Pr
Construct a tadpole in $Z^3$ where (a) the tail pierces the head $k$ times, (b) the head is unknotted and has an edge that is left-most, and (c) the vertex of degree 1 is the unique right-most vertex.  Suppose the head has $m_h$ edges and the tail has $m_t$ edges.  This is a lattice version of something similar to the tadpole sketched in Figure \ref{fig:piercedhead}.  $m_h$ and $m_t$ are now fixed.  Concatenate this tadpole with a polygon with 
$\alpha n - m_h$ edges and with a positive walk with $(1-\alpha)n - m_t$ edges.  This gives the lower bound
$$t^{[k]}(\alpha n, (1-\alpha)n) \ge p_{\alpha n - m_h} c^+_{(1-\alpha)n - m_t}/2$$
and hence $\liminf_{n\to\infty} n^{-1} \log t^{[k]}(\alpha n, (1-\alpha)n) \ge \kappa_3.$  To get an upper bound delete an edge from the tadpole head incident on the vertex of degree 3, to obtain a self-avoiding walk with $n-1$ edges.  Hence 
$\limsup_{n \to \infty} n^{-1} \log t^{[k]}(\alpha n, (1-\alpha)n) \le \kappa_3.$  The first result follows.
For the second result, use the first construction described above but with an unknotted polygon.  This gives:
$$t_{\emptyset}^{[k]}(\alpha n, (1-\alpha)n) \ge p^0_{\alpha n - m_h} c^+_{(1-\alpha)n - m_t}/2$$
and 
$$\liminf_{n \to \infty} n^{-1} \log t_{\emptyset}^{[k]}(\alpha n, (1-\alpha)n) \ge  \alpha \kappa_3^0 + (1-\alpha) \kappa_3.$$
For the upper bound, concatenate an unknotted polygon with $\alpha n$ edges and a self-avoiding walk with $(1-\alpha)n$ edges at each vertex of the polygon, giving:
$$t_{\emptyset}^{[k]}(\alpha n, (1-\alpha)n) \le \alpha n p^0_{\alpha n} c^+_{(1-\alpha)n}$$
and 
$$\limsup _{n \to \infty} n^{-1} \log t_{\emptyset}^{[k]}(\alpha n, (1-\alpha)n) \le  \alpha \kappa_3^0 + (1-\alpha) \kappa_3.$$
This completes the proof.
\qed

\subsection{Tadpoles with a small head}
\label{sec:smallhead} 

If we insist that the head is small, we can obtain stronger results and the behaviour is dominated by the behaviour of the tail.  If we fix the head size and fix the number of piercings of the surface spanned by the head, we get the following for $n$ large:

\begin{theo}
With $h=8$ and $k=1$ fixed, there exist positive numbers $A_1$ and $A_2$ such that
$$ A_1 c_{n-1} (1+o(1)) \le t^{[1]}(8,n-8) \le \sum_m t^{[1]}(m,n-m) \le A_2 c_{n-1}$$
when $n$ is sufficiently large.
\label{theo:smallhead}
\end{theo}
\Pr
Define the pattern 
$$P_3= (0,0,0)-(-1,0,0)-(-2,0,0)-(-2,1,0)-(-2,2,0)-(-1,2,0)-(0,2,0)-(0,1,0)-$$
$$(0,1,-1)-(-1,1,-1)-(-1,1,0)-(-1,1,1)$$
By adding the edge $(0,0,0)-(0,1,0)$ we can convert $P_3$ into a tadpole with a head having eight edges, that is pierced by the tail.  If we consider self-avoiding walks with $n-1$ edges that have $P_3$ as the initial end pattern, Madras' end pattern theorem \cite{Madras} says that
$c_{n-1}(P_3) \ge A_1 c_{n-1} (1+o(1))$.  By adding the edge $(0,0,0)-(0,1,0)$ to convert to a tadpole we have the inequality
$$ A_1 c_{n-1} (1+o(1)) \le t^{[1]}(8,n-8).$$
We can obtain an upper bound by deleting an edge in the head, incident on the vertex of degree 3, to give a self-avoiding walk with $n-1$ edges, and the result follows.
\qed

We can generalize to larger head sizes by concatenating the head of the tadpole with a polygon, and we can construct patterns that generalize from $k=1$ to larger (but fixed) values of $k$.  This extends the statement of the last theorem to give:
\begin{equation}
 A_1 c_{n-1} (1+o(1)) \le t^{[k]}(h,n-h) \le \sum_m t^{[k]}(m,n-m) \le A_2 c_{n-1}
\end{equation}
for fixed $k \ge 1$ and fixed $h >> k$.

In a similar way we can fix the knot type of the head.  For instance, with the head unknotted, $h$ and $k$ fixed with $h >> k$, there are two positive numbers $B_1$ and $B_2$ such that 
\begin{equation}
B_1 c_{n-1} (1+o(1)) \le t_{\emptyset}^{[k]}(h,n-h) \le \sum_m t_{\emptyset}^{[k]}(m,n-m) \le B_2 c_{n-1}.
\end{equation}
The same is true for any fixed knot type.

\section{Other homeomorphism types}
\label{sec:others}

Similar results can be obtained for several other homeomorphism types such as dumbbells, twin tailed tadpoles and two tailed tadpoles, and the proofs are modification of the proofs given in Section \ref{sec:tadpoles}.   Some results will be stated without proof, if the proof is an easy extension or modification of a proof given in Section \ref{sec:tadpoles}.  

\begin{figure}
\centering
\includegraphics[scale=0.3]{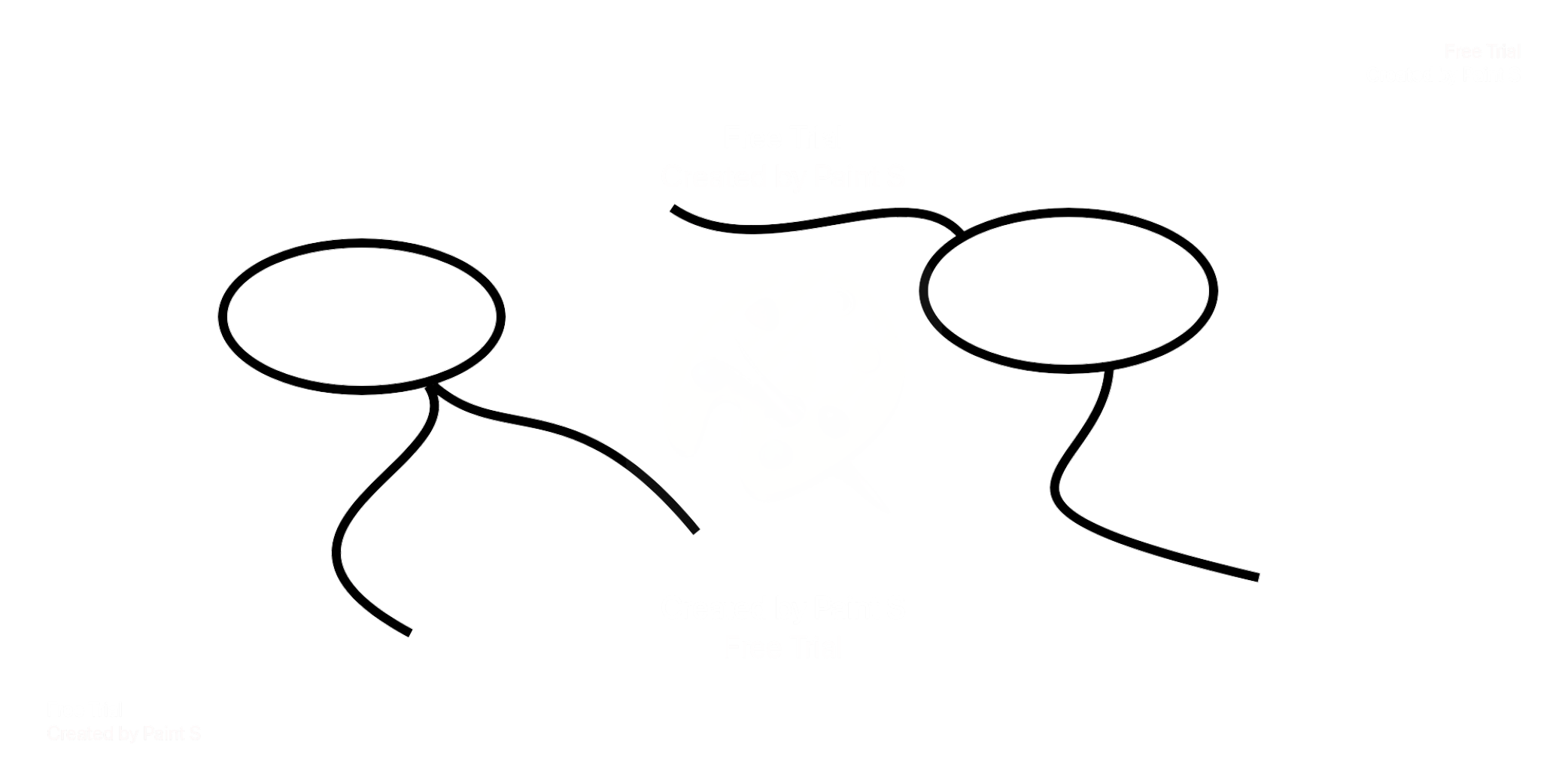}
\caption{Twin tailed tadpole (left) and two tailed tadpole (right).}
\label{fig:tadpoleextensions}
\end{figure}

A dumbbell is a graph with two cycles joined by a sequence of edges.  Dumbbells have two vertices of degree 3 and cyclomatic index 2.  They occur in Sykes' counting theorem \cite{Sykes}.  Twin tailed and two tailed tadpoles \cite{Domb} are sketched in Figure \ref{fig:tadpoleextensions}, and twin tailed tadpoles have been found in protein structures \cite{Sulkowska2019}.


Suppose that $d(h_1, h_2,t)$ is the number of dumbbells with $h_1$ and $h_2$ edges in the two cycles and $t$ edges in the walk between the two cycles.  Then $d_n = \sum_{h_1,h_2} d(h_1, h_2, n-h_1-h_2) $ is the total number of dumbbells with $n$ edges.  We have the following theorem:

\begin{theo}
There are two constants, $A_1$ and $A_2$, such that
$$0 < A_1 \le \liminf_{n \to \infty} \frac{d_n}{c_n} \le \limsup_{n \to \infty} \frac{d_n}{c_n}  \le A_2 \le 1.$$
\end{theo}
\Pr
The proof is a modification of the proof of Theorem \ref{thm:Thm1} but where (for the lower bound) we use two end patterns ($P_0$) at the two ends of a self-avoiding walk of $n-2$ edges, and make use of Madras' end pattern theorem \cite{Madras}.  For the upper bound we delete an edge in each cycle, incident on a vertex of degree 3, to yield a self-avoiding walk with two less edges.
\qed

This shows that dumbbells, like tadpoles, have the same asymptotics as self-avoiding walks.  If the critical exponent $\gamma$ for self-avoiding walks exists, then dumbbells have the same critical exponent.

If the numbers of edges in the two cycles and in the walk all diverge together we have a result exactly analogous to Theorem \ref{theo:divergetogether}.  That is, for any $0 < \alpha_1, \alpha_2 < 1$,
\begin{equation}
\lim_{n \to \infty} n^{-1} \log d(\alpha_1 n, \alpha_2 n, (1-\alpha_1 - \alpha_2)n) = \kappa_d.
\end{equation}

If the surfaces spanned by the two cycles are pierced a fixed number of times, we can prove theorems for dumbbells analogous to those for tadpoles, discussed in Sections \ref{sec:constraints} and \ref{sec:smallhead}.


Twin tailed and two tailed tadpoles are somewhat more difficult.  
Suppose that $a(h,t_1,t_2)$ is the number of twin tailed tadpoles with $h$ edges in the cycle and $t_1$ and $t_2$ edges in the two walks, and $b(h,t_1,t_2)$ is the corresponding count for two tailed tadpoles.  If $a_n = \sum_{t_1,t_2} a(n-t_1-t_2, t_1, t_2)$, the total number of $n$-edge twin tailed tadpoles, and $b_n$ is the total number of $n$-edge two tailed tadpoles, it is easy to prove that the total numbers of both homeomorphism types grow at the same exponential rate as self-avoiding walks.  A pattern theorem argument \cite{Kesten} can be used to prove the following:
\begin{theo}
For $n$ sufficiently large
$$a_n \ge \epsilon_a n c_n \left(1- e^{-\beta_a n} \right)$$
for some $\epsilon_a, \beta_a > 0$,
so the number of twin tailed tadpoles grows at least $n$ times as fast as the number of self-avoiding walks. 
\end{theo}
\Pr
In $Z^2$ define $P_3$ to be the pattern:
$$P_3 = (0,0)-(0,1)-(-1,1)-(-1,2)-(-1,3)-(0,3)-(0,2)-(1,2).$$
In higher dimensions, let $P_3$ be the same pattern occurring in the $(x_1,x_2)$-plane.  $P_3$ can be modified to form a twin tailed tadpole with a cycle of size 4 by deleting $(0,1)-(-1,1)-(-1,2)$ and adding $(0,1)-(0,2)-(-1,2)$
By Kesten's pattern theorem \cite{Kesten} $P_3$ occurs at least $\epsilon_a n$ times on all except exponentially few self-avoiding walks, for some positive $\epsilon_a$.  Choose one of these occurrences (in $\epsilon_a n$ ways) and convert to a twin tailed tadpole by adding and deleting edges as above,  The result follows.
\qed

For two tailed tadpoles one can use a similar argument (using the pattern $P_0$) to show that
\begin{equation}
b_n \ge \epsilon_b (n-1) c_{n-1} \left(1- e^{-\beta_b (n-1)}\right)
\end{equation}
for some $\epsilon_b, \beta_b > 0$.

If the numbers of edges in the cycle and in the two walks all diverge together, it is easy to obtain information about the exponential growth rate.    Then
\begin{equation}
\lim_{n \to \infty} n^{-1} \log a((1-\alpha_1 -\alpha_2)n, \alpha_1 n, \alpha_2 n) = 
\lim_{n \to \infty} n^{-1} \log b((1-\alpha_1 -\alpha_2)n, \alpha_1 n, \alpha_2 n) = \kappa_d
\end{equation}
for all $0 < \alpha_1, \alpha_2 < 1$.  In three dimensions the cycle can be knotted and we can derive corresponding results when the knot type is fixed.  To give an example, if the cycle is unknotted then
\begin{equation}
\lim_{n \to \infty} n^{-1} \log a_{\emptyset}((1-\alpha_1 -\alpha_2)n, \alpha_1 n, \alpha_2 n) = 
(1-\alpha_1-\alpha_2) \kappa_3^0 + (\alpha_1 + \alpha_2) \kappa_3
\end{equation}
with a similar result for $b_{\emptyset}((1-\alpha_1 -\alpha_2)n, \alpha_1 n, \alpha_2 n)$.

\section{Discussion}
\label{sec:discussion}

We have proved several results about the growth rate of tadpoles in a lattice including cases where the tail and head grow together or where the head is small compared to the tail.  We have also considered the effects of various constraints such as insisting that the head has a particular knot type, or requiring the tail to pierce the surface spanned by the head a specified number of times.  Some of these results are relevant to tadpole architectures found in protein structures \cite{Sulkowska2024,Sulkowska2019}, and several of our results have been observed numerically \cite{Beaton}.  We have given some similar results for other homeomorphism types such as dumbbells, twin tailed tadpoles and two tailed tadpoles.

\section*{Acknowledgement  }
The author would like to thank Chris Soteros, Joanna Sulkowska, Nick Beaton and Buks van Rensburg for helpful discussions.


\end{document}